\documentclass
[amsfonts,10pt,nofootinbib,floats,superbib,prl,twocolumn,letterpaper,tightenlines,preprint,showpacs,bibnotes,titlepage,noshowkeys,hyperref]{revtex4}%
\usepackage{amsfonts}
\usepackage{amsmath}
\usepackage{amssymb}
\usepackage{graphicx}
\usepackage{alltt}%
\providecommand{\U}[1]{\protect\rule{.1in}{.1in}}

\begin{document}
\preprint{arXiv}
\title[Huygens synchronization]{Huygens synchronization of two pendulum clocks}
\author{Henrique M. Oliveira$^{1,3}$}
\email{holiv@math.tecnico.ulisboa.pt}
\author{Lu\'{\i}s V. Melo$^{2,3}$}
\affiliation{$^{1}$Center of Mathematical Analysis Geometry and Dynamical Systems, $^{2}%
$INESC-MN and IN, Av. Alves Redol 9, 1000-029 Lisboa, ~$^{3}$Instituto
Superior T\'{e}cnico, Universidade de Lisboa, Av. Rovisco Pais, 1049-001
Lisboa, Portugal}
\keywords{Synchronization, Huygens' Clocks, Phase Locking}
\pacs{05.45.Xt}

\begin{abstract}
The synchronization of two pendulum clocks hanging from a wall was first
observed by Huygens during the XVII century. This type of synchronization is
observed in other areas, and is fundamentally different from the problem of
two clocks hanging from a moveable base. We present a model explaining the
phase opposition synchronization of two pendulum clocks in those conditions.
The predicted behavior is observed experimentally, validating the model.

\end{abstract}
\volumeyear{ }
\volumenumber{ }
\issuenumber{ }
\eid{ }
\startpage{1}
\endpage{ }
\maketitle

\textit{Introduction.-- }The synchronization between two periodic systems
connected through some form of coupling is a recurrent, and still pertinent,
problem in Nature, and in particular in Physics. During the 17th century
Huygens, the inventor of the pendulum clock, observed phase or phase
opposition coupling between two heavy pendulum clocks hanging either from a
house beam and later from a board sitting on two chairs \cite{Huy}. These two
systems are inherently different in terms of the coupling process, and in
consequence of the underlying model. The later case has been thoroughly
studied \cite{Benn,Col,Col2,Frad, Jova,Martens,Oud,Sen} by considering
momentum conservation in the clocks-beam system. The first case has been
approached in a few theoretical works \cite{Abr, Abr2,Nun, Vass}. We present a
mathematical model where the coupling is assumed to be attained through the
exchange of impacts between the oscillators (clocks). This model presents the
additional advantage of being independent of the physical nature of the
oscillators, and thus can be used in other oscillator systems where
synchronization and phase locking has been observed \cite{Pit}.The model
presented starts from the Andronov \cite{And} model of the phase-space limit
cycle of isolated pendulum clocks and assumes the exchange of single
\textit{impacts} (sound solitons, for this system) between the two clocks at a
specific point of the limit cycle. Two coupling states are obtained, near
phase and near phase opposition, the latter being stable. Our experimental
data, obtained using a pair of similar pendulum clocks hanging from an
aluminum rail fixed to a masonry wall, match the theoretical predictions and simulations.

\textit{Andronov model.}-- The model for the isolated pendulum
clock\ (\textit{Andronov clock) }has been studied using models with viscous
friction by physicists \cite{Benn,Col,Frad, Jova,Martens,Oud,Sen}. However,
Russian mathematicians lead by Andronov published a work \cite{And} where the
stability of the model with dry friction is established. The authors prove the
existence and stability of the limit cycle.

\begin{figure}
[h]
\begin{center}
\includegraphics[
height=2.1075in,
width=2.2753in
]%
{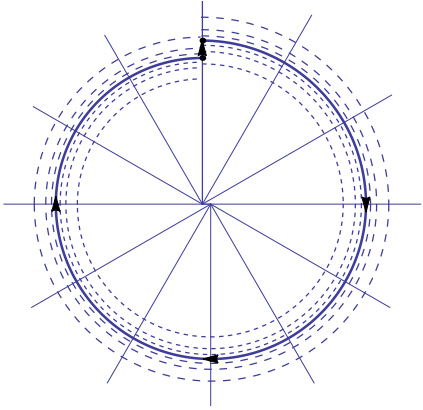}%
\caption{Limit cycle of an isolated clock represented as a solid curve in the
phase space. Horizontal axis represents the angular position and in the
vertical axis the velocity. We use normalized coordinates to get arcs of
circles.}%
\label{Fig1}%
\end{center}
\end{figure}

We adopt as basis for our work the aforementioned model, assuming that dry
friction predominates. Using the normalized angular coordinate $q$, the
differential equation governing the pendulum clock is%
\begin{equation}
\ddot{q}+\mu\operatorname{sign}\dot{q}+\omega^{2}q=0, \label{Andronov}%
\end{equation}
where $\mu>0$ is the dry friction coefficient, $\omega$ is the natural angular
frequency of the pendulum and $\operatorname{sign}\left(  x\right)  $ a
function giving $-1$ for $x<0$ and $+1$ otherwise. In \cite{And} was
considered that, in each cycle, a fixed amount of normalized kinetic energy
$\frac{h^{2}}{2}$ is given by the escape mechanism to the pendulum to
compensate the loss of kinetic energy due to dry friction in each complete
cycle. We call to the transfer of kinetic energy a \textit{kick}. We set the
origin such that the kick is given when $q=-\frac{\mu}{\omega^{2}}$, which is
very close to $0$. The phase portrait is shown in Fig. \ref{Fig1}.

We consider initial conditions $q\left(  t=0\right)  =-\frac{\mu}{\omega^{2}}$
and $\dot{q}\left(  t=0\right)  =v_{0}$. We draw a Poincar\'{e} section
(\cite{Bir} vol. II, page 268) as the half line $q=-\frac{\mu}{\omega^{2}}%
^{+}$ and $\dot{q}>0$ \cite{And}. The symbol $+$ refers to the fact that we
are considering that the section is taken immediately after the kick. Solving
the differential equations (\ref{Andronov}) sectionally we notice that there
is a loss of velocity $-\frac{4\mu}{\omega}$ due to friction during a complete
cycle. Considering $v_{n}=\dot{q}\left(  \frac{2n\pi}{\omega}^{+}\right)  $
the velocity at the Poincar\'{e} section in each cycle one obtains the
discrete dynamical system%
\begin{equation}
v_{n+1}=\sqrt{\left(  v_{n}-\frac{4\mu}{\omega}\right)  ^{2}+h^{2}}\text{,}
\label{recu}%
\end{equation}
which has the asymptotically stable\ fixed point \cite{And}%
\[
v_{f}=\frac{h^{2}\omega}{8\mu}+\frac{2\mu}{\omega}\text{.}%
\]
The fixed point (\ref{recu})\ attracts initial conditions $v_{0}$\ in the
interval $\left(  \frac{4\mu}{\omega^{2}},+\infty\right)  $.

\textit{Model for two pendulum clocks.}-- We consider two pendulum clocks
suspended at the same wall. When one clock receives the kick, the impact
propagates in the wall and the second clock is slightly perturbed by a
traveling wave. The perturbation is assumed instantaneous since the time of
travel of sound in the wall between the clocks is assumed very small compared
to the period. The interaction was studied geometrically and qualitatively by
Abraham \cite{Abr2, Abr} without the heavy computations of analytical
treatments. However, that approach does not give estimates on the speed of convergence.

In Vassalo-Pereira\cite{Vass}, the theoretical problem of the phase locking is
tackled. The author makes the assumptions:

\begin{enumerate}
\item dry friction,

\item the pendulums have the same exact natural frequency $\omega$,

\item the perturbation in the momentum is always in the same vertical
direction in the phase space, see also \cite{Abr2, Abr}.

\item The perturbation imposes a discontinuity in the momentum but not a
discontinuity in the dynamic variable. The interaction between clocks takes
the form of a Fourier series \cite{Vass}.
\end{enumerate}

Vassalo-Pereira deduced that the two clocks synchronize with zero phase
difference. This is the exact opposite of Huygens first remarks \cite{Huy} and
our experimental observations, where phase opposition was observed. Therefore,
we propose a modified model accounting for a difference in frequency between
the two clocks.

Consider two oscillators indexed by $i=1,2$, $\phi_{i}\left(  t\right)  $ is
the time difference of clock $i$ relative to the other. Each oscillator
satisfies the differential equation%
\begin{equation}
\ddot{q}_{i}+\mu_{i}\operatorname{sign}\dot{q}_{i}+\omega_{i}^{2}q_{i}%
=-\alpha_{i}\delta\left(  t-\phi_{i}\left(  t\right)  \right)  ,\text{ for
}i=1,2\text{,} \label{coupledandro}%
\end{equation}
when $q_{i}=-\frac{\mu_{i}}{\omega_{i}^{2}}$, the kinetic energy of each
oscillator is increased by the fixed amount $h_{i}$ as in the \textit{Andronov
model}. The coupling term is the normalized force $\alpha_{i}\delta\left(
t-\phi_{i}\left(  t\right)  \right)  $, where $\delta$ is the Dirac delta
distribution and $\alpha_{i}$ a constant with acceleration dimensions.

The sectional solutions of the differential equation \ref{coupledandro} are
obtainable when the clocks do not suffer kicks. To treat the effect of the
kicks we construct a discrete dynamical system for the phase difference. The
idea is similar to the construction of a Poincar\'{e} section. If there exists
an attracting fixed point for that dynamical system, the phase locking occurs.

Our assumptions are

\begin{enumerate}
\item Dry friction.

\item The pendulums have natural angular frequencies $\omega_{1}$ and
$\omega_{2}$ near each other with $\omega_{1}=\omega+\varepsilon$ and
$\omega_{2}=\omega-\varepsilon$, where $\varepsilon\geq0$ is a small
parameter, typically $\varepsilon<10^{-3}$.

\item The perturbation in the momentum is always in the same vertical
direction in the phase space\cite{Abr2, Abr}.

\item Since the clocks have the same construction, the energy dissipated at
each cycle of the two clocks is the same, $h_{1}=h_{2}=h<2\times10^{-2}$. The
friction coefficient is the same for both clocks, $\mu_{1}=\mu_{2}=\mu
<4\times10^{-4}$.

\item The kick is instantaneous. This is a reasonable assumption, since in
general the perturbation propagation time between the two clocks is several
orders of magnitude lower than the periods.

\item The interaction is symmetric, the coupling has the same constant
$\alpha$ when the clock $1$ acts on clock $2$ and conversely. In our model we
assume that $\alpha$ is very small.
\end{enumerate}

All values throughout the paper are in SI units.

In this paper the function $\arctan$ when notated with two variables is the
generalization of the usual $\arctan$ to incorporate the information about the
quadrant. Thus, we define%
\[
\arctan\left(  y,x\right)  =\left\{
\begin{array}
[c]{cccc}%
\arctan\frac{y}{x} & \text{ if } & x\geq0 & y\geq0,\\
\pi+\arctan\frac{y}{x} & \text{ if } & x<0, & \\
2\pi+\arctan\frac{y}{x} & \text{ if } & x\geq0 & y<0.
\end{array}
\right.
\]

The phase of each clock at the limit cycle is defined as%
\[
\Phi_{i}=\left\{
\begin{array}
[c]{c}%
\arctan\left(  \omega_{i}q_{i}\left(  t\right)  +\frac{\mu}{\omega_{i}}%
,\dot{q}_{i}\right)  \text{ if }\dot{q}_{i}\geq0\text{,}\\
\arctan\left(  \omega_{i}q_{i}\left(  t\right)  -\frac{\mu}{\omega_{i}}%
,\dot{q}_{i}\right)  \text{ if }\dot{q}_{i}<0\text{.}%
\end{array}
\right.  \text{ }i=1,2\text{.}%
\]
If there is no interaction between the two clocks, the phase of each clock is%
\[
\Phi_{i}=\arctan\left(  \sin\omega_{i}t,\cos\omega_{i}t\right)  =\omega
_{i}t\text{, }i=1,2\text{.}%
\]

For the purpose of analyzing the asymptotic properties of the interacting
system, and without loss of generality, we consider that at initial time
oscillator $1$ and $2$ are isolated, both at each limit cycle. The phase
difference is $\phi\left(  t\right)  =\Phi_{2}\left(  t\right)  -\Phi
_{1}\left(  t\right)  $, between the two Andronov clocks. When the two clocks
are isolated from each other, and considering the period of the fastest
$T=\frac{2\pi}{\omega+\varepsilon}$ the phase difference can be seen as a map
from the circle $S^{1}$ in itself with solution%
\begin{equation}
\phi_{n}\equiv4\pi n\frac{-\varepsilon}{\omega+\varepsilon}+\phi_{0}\text{
}\left(  \operatorname{mod}2\pi\right)  \text{,} \label{discretephase}%
\end{equation}
which is a rigid rotation of the circle where the discrete variable $n$
denotes the number of cycles of the fastest clock with natural frequency
$\omega_{1}$.

When the coupling is established we have to modify \ref{discretephase} to
obtain the phase difference of the coupled system $\varphi_{n}$. The
asymptotic properties of the discrete function $\varphi_{n}$ determine the
limit properties of the phase difference of the coupled system.

The notation is simplified if we consider the function $\gamma\left(
\varphi\right)  $ such that%
\[
\gamma\left(  x,y\right)  =\left\{
\begin{array}
[c]{c}%
\frac{h^{2}y}{8\mu}+\frac{2\mu}{y}\text{, if }0\leq x<\frac{\pi}{2}\text{,}\\
\frac{h^{2}y}{8\mu}\text{, if }\frac{\pi}{2}\leq x<\frac{3\pi}{2}\text{,}\\
\frac{h^{2}y}{8\mu}-\frac{2\mu}{y}\text{, if }\frac{3\pi}{2}\leq
x<2\pi\text{,}%
\end{array}
\right.
\]

We assume that the natural frequencies are close. A relatively large
difference of $28$ seconds per day in the movement of the clocks with natural
periods in the order of $1.42%
\operatorname{s}%
$, implies that $\varepsilon\sim10^{-3}$.

\begin{figure}
[h]
\begin{center}
\includegraphics[
trim=0.000000in 0.000000in -0.062730in 0.000000in,
height=1.7002in,
width=3.2638in
]%
{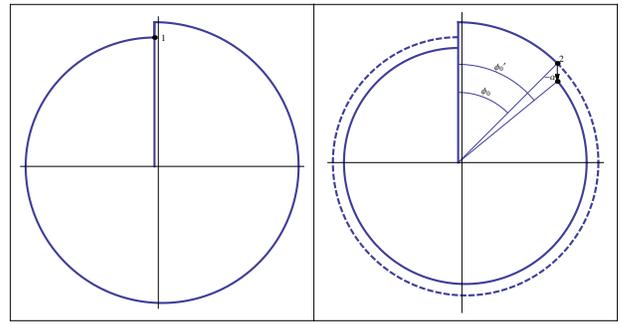}%
\caption{Interaction of clock $1$ on clock $2$ at $t=0^{+}$. We see the
original limit cycle, before interaction, and the new one in solid and the
original limit cycle in dashed. Note that the value of $\alpha$ and of $h$ are
greatly exaggerated to provide a clear view. The effect of the perturbation is
secular and cumulative.}%
\label{Fig2}%
\end{center}
\end{figure}

\begin{figure}
[h]
\begin{center}
\includegraphics[
height=1.7288in,
width=3.2949in
]%
{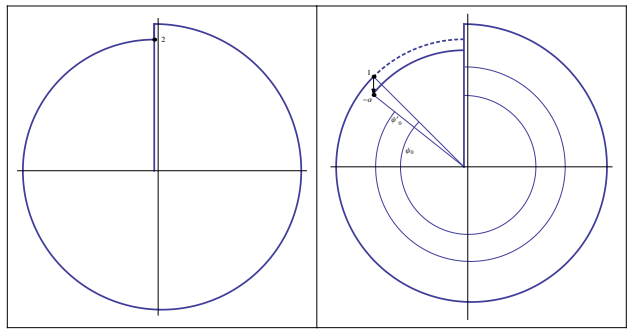}%
\caption{Second interaction. Interaction of clock $2$ on clock $1$ when clock
$2$ reaches its impact position. All the features are similar to the Fig.
\ref{Fig2}. }%
\label{Ultima}%
\end{center}
\end{figure}

This means that each clock will receive a perturbative kick from the other per
cycle of the fastest one. This assumption would fail once in $n=10^{4}$ cycles
of the fastest clock. Suppose that the clocks are brought to interaction at
$t_{0}=0$. The fastest clock (number $1)$ is at initial position
\[
q_{1}\left(  0^{-}\right)  =-\frac{\mu}{\omega+\varepsilon}\text{, }\dot
{q}_{1}\left(  0^{-}\right)  =\frac{h^{2}\omega+\varepsilon}{8\mu}-\frac{2\mu
}{\omega+\varepsilon}.
\]
Solving sectionally the differential equations with the two small
interactions, we can construct a discrete dynamical system taking into account
the two interactions per cycle seen in figure \ref{Fig2} and \ref{Ultima}.
After that, we compute the phase difference when clock $1$ returns to the
initial position.%

%

Expanding the solutions of the differential equations and computing the phase
difference we get in first order of $\alpha$ and $\varepsilon$ the iterative
scheme%
\[
\phi_{n+1}=\Xi\left(  \phi_{n}\right)  =\phi_{n}+4\pi\frac{-\varepsilon
}{\omega+\varepsilon}+\frac{2\alpha\sin\phi_{0}}{\gamma\left(  \phi_{0}%
,\omega\right)  }+h.o.t,
\]
The iteration of the map $\Xi\left(  x\right)  $, defines the asymptotic
properties of the phase difference. There are two fixed points of $\Xi$ in the
interval $\left[  0,2\pi\right]  $. We have the first order approximation for
the fixed point $\phi_{f}$%
\[
\sin\phi_{f}=\frac{\pi h^{2}\varepsilon}{4\alpha\mu},
\]
with two solutions, respectively stable and unstable%
\[
\phi_{f}^{s}=\pi-\arcsin\frac{\pi h^{2}\varepsilon}{4\alpha\mu}\text{, }%
\phi_{f}^{u}=\arcsin\frac{\pi h^{2}\varepsilon}{4\alpha\mu}\text{.}%
\]
The derivative of $\Xi$ at the fixed point must be $\left\vert \Xi^{\prime
}\left(  x_{f}\right)  \right\vert <1$ to have stability. The stability
condition at $\phi_{f}^{s}\in\left[  \frac{\pi}{2},\pi\right]  $ and the
condition about the argument of the function $\arcsin$ gives%
\begin{equation}
\frac{\pi h^{2}}{4\mu}\varepsilon<\alpha<\frac{h^{2}\omega}{8\mu}\text{, i.e.,
}\frac{h^{2}}{8\mu}\pi\Delta\omega<\alpha<\frac{h^{2}}{8\mu}\overline{\omega
}\text{.} \label{Conditions}%
\end{equation}
The limit of the phase difference is, in first order, $\pi-\arcsin\frac{\pi
h^{2}\varepsilon}{4\alpha\mu}$. When the system reaches this limit the
corrections of phase are null for both clocks.

If the initial conditions are very near, with clock $1$ slightly behind clock
$2$, one can have one overtaking but, after that one there are no more and the
phase difference of the clocks tends to the same limit $\pi-\arcsin\frac{\pi
h^{2}\varepsilon}{4\alpha\mu}$. The asymptotic behavior of the phenomenon is
exactly the same.

\textit{Simulation}.-- To study the Huygens synchronization we used numerical
simulations. We applied the map $\Xi\left(  x\right)  $ without performing the
Taylor expansion. We used the environment of Wolfram Mathematica 9.0 to
produce the computations. The values of $\mu$, $h$, $\omega$, $t_{0}$ were
taken realistically from the experimental setup and kept fixed throughout the
simulations. The coupling constant $\alpha$ and the half-difference between
the clocks frequencies $\varepsilon$\ are adjusted in simulations.

Additionally, we introduced noise in the model with normal distribution acting
directly on the phase. The effect of noise is to mimic the small perturbations
that occur in the lab, e.g., vibrations in the wall and the stochastic changes
of the level of the interaction, cycle after cycle. The strength of this
stochastic effect is given by the parameter $\rho$. When the noise function is
not used, i.e. $\rho=0$, if the parameters are in the convergence region given
by conditions (\ref{Conditions}) we have a fixed convergence point of $\Omega$.

\textit{Experimental.}-- Experiments were setup using a standard optical rail
(Eurofysica) rigidly attached to a wall to which two similar clocks were fixed
through an optical rain modified rail carriers. The sound propagation speed in
Al is $6420%
\operatorname{m}%
\operatorname{s}%
^{-1}$, leading to a propagation time of the order of $3.0\times10^{-5}%
\operatorname{s}%
$, which is negligible in face of the $\ \symbol{126}1.4%
\operatorname{s}%
$ period of the oscillators. Other materials we tried (MDF and fiberglass) but
no coupling could be observed. The mass-driven pendulum clocks used were
Acctim 26268 Hatahaway. The chime mechanism was inhibited in order to reduce
mechanical noise. Time was measured using one U-shaped LED emitter-receptor
TCST 1103, connected to a Velleman K8055 USB data acquisition board operated
with custom-developed software running in a standard personal computer (PC).
The uncertainty in time acquisition typically expected in the $%
\operatorname{ms}%
$ range for the PC was overcame by performing running averaging of the period
data, up to $1000$. The data files were then processed offline using
Mathematica. In order to obtain the appropriate parameters for the simulation,
the pendulums were filmed and the movement quantified using free software
\textit{Tracker4.84} from OSP (https://www.cabrillo.edu/\symbol{126}dbrown/tracker/).

\textit{Results and discussion.}-- Observing the movement of the pendulum
alone (as a damped oscillator; initially at the limit cycle), we noticed a
decrease of the maximum velocity of the pendulum according to a linear fit
$v_{\max}=0.2228-0.0023n$, where $n$ is the number of cycles, with correlation
coefficient $0.994$. The decrease of velocity per cycle predicted by the
Andronov model is $\frac{4\mu}{\omega}$. With the value of the period $T=1.4$
and $\omega=\frac{2\pi}{T}=4.48799$, we can estimate the value of $\mu
=\frac{\omega\Delta v}{4}\approx2.54\times10^{-3}$.

The value of $h$ can also be easily established by studying the movement at
the limit cycle. We then have the maximum velocity $v_{f}=\frac{h^{2}\omega
}{8\mu}+\frac{2\mu}{\omega}$. We found consistently that the maximum velocity
at the limit cycle is $v_{f}=0.223$. Therefore, $h\approx0.032$.

We used different possible values of natural frequency difference. For the
average natural angular frequency we take the same value of $T=1.4$, hence
$\omega=4.48799=2\pi/T$. The fastest clock has a natural frequency of
$\omega+\varepsilon$ and the slowest $\omega-\varepsilon$. The angular
frequency difference is in first order $\varepsilon=\frac{2\pi}{T^{2}}\Delta
t$, where $\Delta t$ is half of the period difference. Notice that when
$\Delta t=2\times10^{-4}$, with a delay between the clocks of $24.6$ $%
\operatorname{s}%
$ per day for the non coupled pair of clocks, the value of $\varepsilon$ is
$\varepsilon=6.4\times10^{-4}$. We used values of $\varepsilon$ in the range
$10^{-4}$ to $10^{-3}$ as a realistic estimate for the performance of our setup.

\begin{figure}
[h]
\begin{center}
\includegraphics[
height=2.2576in,
width=3.3818in
]%
{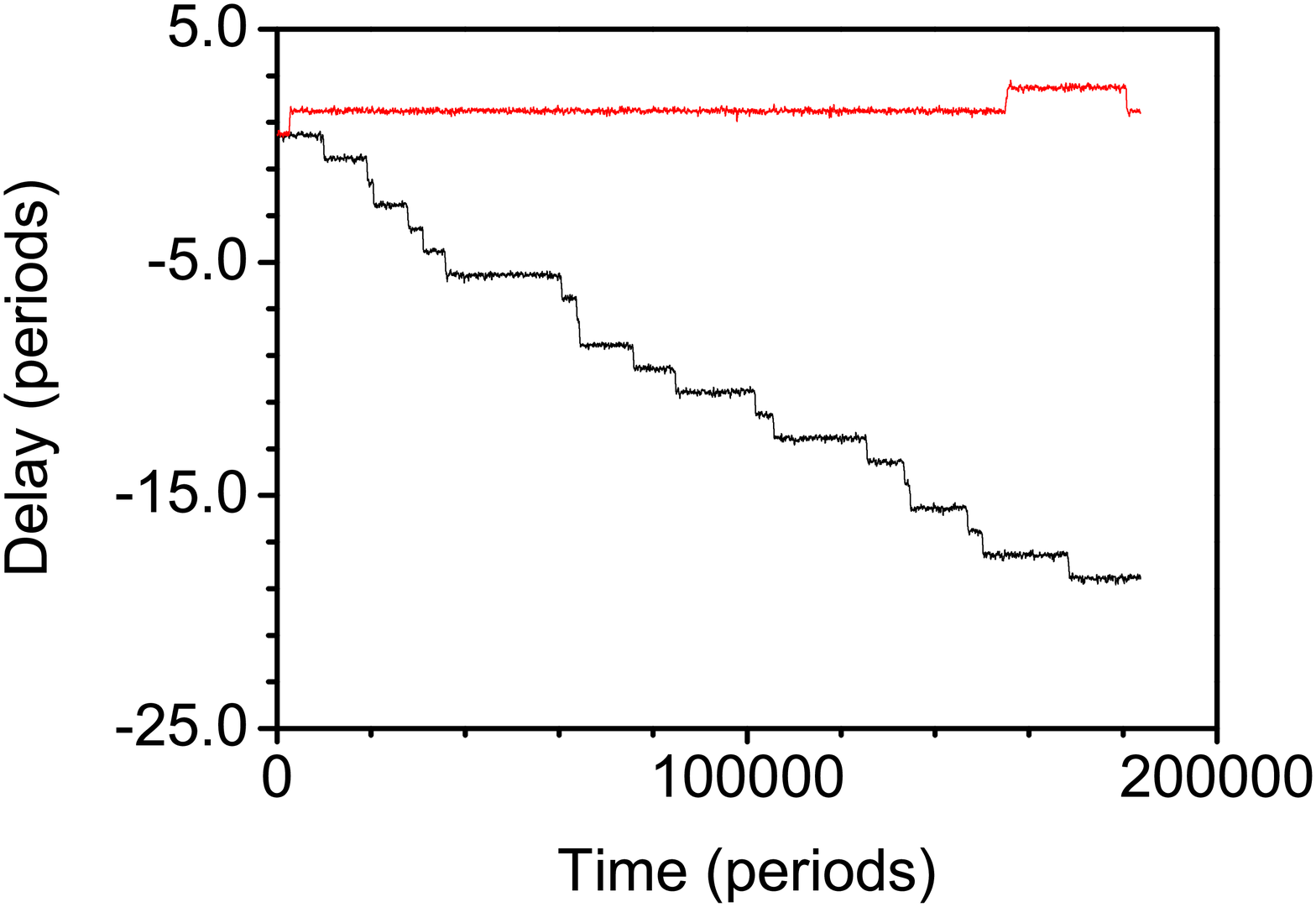}%
\caption{Simulations of delay between the two clocks in period units for two
frequency differences. Upper curve: $\ \epsilon=1.5\times10^{-4}%
\operatorname{s}^{-1}$. Lower curve: $\ \epsilon=3.0\times10^{-3}%
\operatorname{s}^{-1}.$}%
\label{Fig4}%
\end{center}
\end{figure}

The fixed parameters used for the simulations are then $\mu=2.54\times10^{-3}%
$, $h=0.032$, $\omega=4.48799$, $t_{0}=0.8\pi$, and the phenomenological noise
coefficient of $\rho=0.093$, which fits the ripple observed at the
experimental data. When we choose $\varepsilon=3\times10^{-3}$, corresponding
to a \ natural delay of $116%
\operatorname{s}%
$ per day with the clocks in the isolated state, a value of $\alpha
=7\times10^{-4}$ yielded results matching the experimental data. We expect
more frequent escapes from stable states than if we choose $\varepsilon
=1.5\times10^{-4}$, corresponding to a natural delay of $2.9%
\operatorname{s}%
$ per day. These values correspond to the values that could realistically be
obtained in our experimental setup. The plots can be seen in Fig. \ref{Fig4}.%

Notice the small differences assumed for the frequencies, of the same order of
the values observed for independent clocks. The time difference stabilizes in
horizontal plateaus, corresponding to phase opposition coupling. The
stochastic term introduced in the simulation unsets the system at some point,
and then the phase difference increases quickly as the fastest clock runs away
until the next synchronization plateau is reached, one or sometimes two
periods away. For the simulation with the smaller difference between
frequencies, the number of transitions between plateaus is smaller, as
expected since the stability is much easier to reach and maintain.

\begin{figure}
[h]
\begin{center}
\includegraphics[
height=2.5019in,
width=3.5319in
]%
{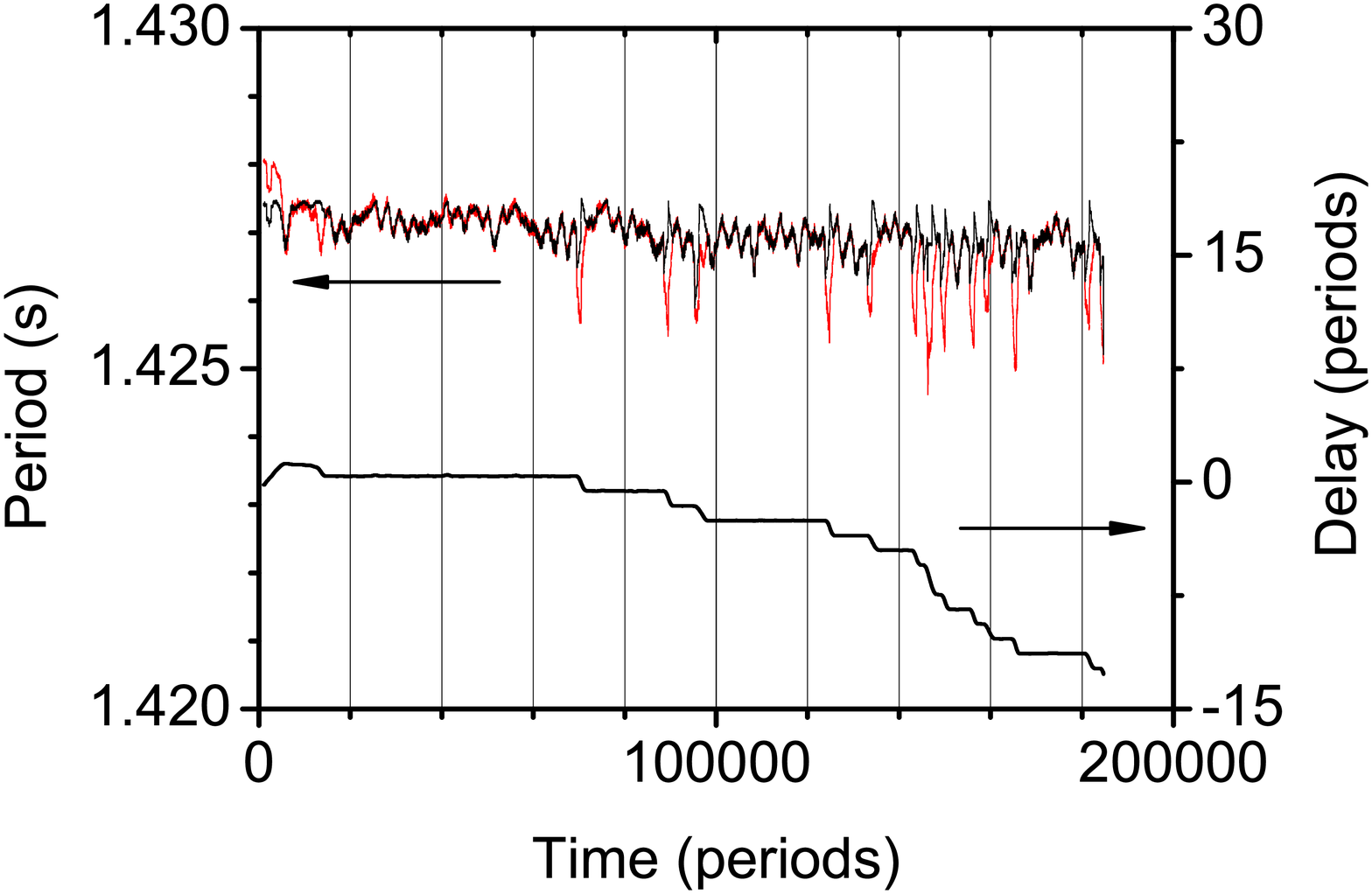}%
\caption{Phase difference between the clocks in period units over more than
three days (lower curve, right axis) and periods of the two clocks (upper
curves, left axis; lighter for clock $\#1$, darker for clock $\#2$). The
initial behavior corresponds to mechanical stabilization of the clocks during
the firs few hours of the experiment.}%
\label{Fig5}%
\end{center}
\end{figure}

This is strikingly similar to the behavior observed in Fig. \ref{Fig5} for the
actual clocks (right axis). The number of synchronization plateaus is of the
same order and can be fine-tuned using the stochastic parameter in the
simulation. It was observed that the system could be unsettled by a number of
external noise sources, e.g. from doors closing nearby in the building, people
entering or leaving the room, or even the elevator stopping, than proceeding
to the next synchronization plateau.%

The periods of both clocks from the same experiment can be seen on the same
figure (left axis). The periods vary together within an interval of about $1%
\operatorname{ms}%
$ around $1.427%
\operatorname{s}%
$ when the clocks are coupled with correlation coefficient above $0.97$ in the
coupled state. Notice the almost perfect coincidence of the two curves except
when the system leaves coupled states. Notice also the unstability of the
coupled period, varying over an interval of almost $1%
\operatorname{ms}%
$. When coupling is lost the period of one clock decreases sharply (up to $2%
\operatorname{ms}%
$ or more) and the period of the other clock increases by a (smaller) amount.
These perturbations in the periods are coincident with the loss of phase
opposition coupling. Although one may expect changes in the frequency even
when the clocks are not coupled, due to the interaction between them, the
difference in period can be estimated at around $2%
\operatorname{ms}%
$, corresponding to a difference in frequency of the order of $6\times10^{-3}%
\operatorname{s}%
^{-1}$. This asymmetry of the coupled period relative to the original periods
is predicted by the model. Both periods return to the previous baseline value
when the coupling is restored.
\begin{figure}
[h]
\begin{center}
\includegraphics[
height=2.3549in,
width=3.5769in
]%
{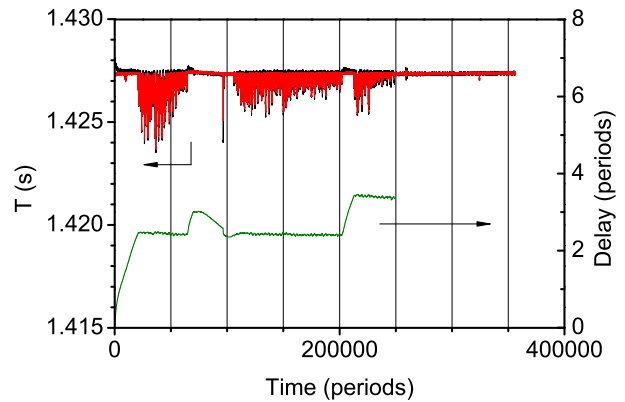}%
\caption{Phase difference between the clocks in period units over more than
three days (lower curve, right axis) and periods of the two clocks (upper
curves, left axis; lighter for clock $\#1$, darker for clock $\#2$). The
initial behavior corresponds to mechanical stabilization of the clocks during
the firs few hours of the experiment. After running for around $250000$
periods, clock $\#2$ is stopped, and clock $\#1$ keeps running undisturbed.}%
\label{Fig6}%
\end{center}
\end{figure}

Fig. \ref{Fig6} shows data for another experiment. In this case the free clock
frequencies were closer. Both periods are remarkably coincident, but vary in
an interval in excess of $10%
\operatorname{ms}%
$, when the clocks are coupled. Since the frequencies are closer, the
synchronization should be easier to maintain, hence the low number of
plateaus, but also should be slower to attain, hence the longer transitions
between plateaus. If the perturbation is large eneough, especially if the
frequencies are very close, it is possible to attain plateaus both above and
below. Between $t=100000T$ and $t=112000T$ approximately the synchronization
is lost, and the periods become separated by more than $100%
\operatorname{\mu s}%
$ (corresponding to frequency difference around $3\times10^{-4}%
\operatorname{s}%
^{-1}$), but become stable (within $1%
\operatorname{\mu s}%
$). At instant $t=148000T$ approximately clock $\#2$ is stopped. From that
moment on the period of the remaining working clock becomes stable within
about $10%
\operatorname{\mu s}%
$, an interval one order of magnitude below. This confirms that the clocks
strongly disturb one another, but also that both periods are kept at the same
value in order to keep the synchronization at the expense of some frequency unstability.

\textit{Conclusions.}-- We have developed a model explaining the Huygens
problem of synchronization between two clocks hanging from a wall. In this
model each clock transmits once per cycle a sound pulse that is translated in
a pendulum speed change. An equilibrium situation is obtained for almost
half-cycle phase difference. These predictions match remarkably the
experimental data obtained for two similar clocks hanging from a wall.

\textbf{Acknowledgement} This work was partially funded by FCT/Portugal
through projects PEst-OE/EEI/LA0009/2013 for CMAGDS and
PEst-OE/CTM/LA0024/2013 for INESC-MN and IN.

\bibliographystyle{apsrev}
\bibliography{BibloH2}

\end{document}